# IoT-Fog: A Blockchain-based Middleware Framework for Communication Security in the Internet of Things

**Tanweer Alam**
Department of Computer Science
Faculty of Computer and Information Systems
Islamic University of Madinah, Saudi Arabia
Email: tanweer03@iu.edu.sa

**Abstract**

The fog computing is the emerging technology to compute, store, control and connecting smart devices with each other using cloud computing. The Internet of Things (IoT) is an architecture of uniquely identified interrelated physical things, these physical things are able to communicate with each other and can transmit and receive information. This research presents a framework of the combination of the Internet of Things (IoT) and Fog computing. The blockchain is also the emerging technology that provides a hyper, distributed, public, authentic ledger to record the transactions. Blockchains technology is a secured technology that can be a boon for the next generation computing. The combination of fog, blockchains, and IoT creates a new opportunity in this area. In this research, the author presents a middleware framework based on the blockchain, fog, and IoT. The framework is implemented and tested. The results are found positive.

**1.0 Introduction:**

The blockchains technology (BT) and IoT are pondered emerging approaches that can change our lives in the next generation computing. The BT is much secured and authentic than other technologies. It can be optimistic to provide us high trust in the secure transactions in the heterogeneous network where everything with connecting and talks to each other [1]. Fog computing can play an important role in IoT and BT network. It can compute, store and control the transaction. The IoT is the ongoing research to connect the physical objects, machines and embedded devices to wireless communication and the internet. According to the Statista report, the IoT connected devices will reach more than 75 billion up to 2025 [2]. This report shows us the importance of the proposed framework.

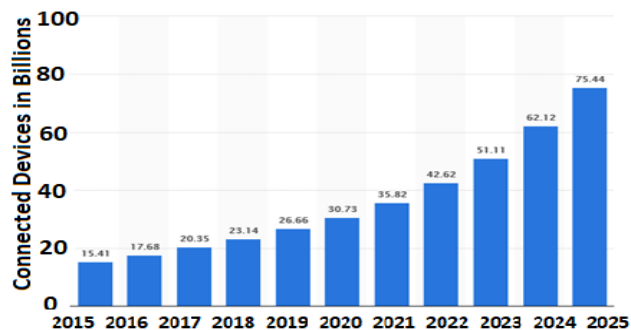

Figure 1. Statista report of connecting devices between 2015-2025 [1]

The blockchain is considered as validating and selecting a chain of blocks. Every IoT node is able to perform validation and selection criteria. The transactions fired by the IoT node are converted into blocks. The blocks are transmitted to the public network and this block will reach every IoT node but this block will be selected by the valid node and verify the hash code. The hyper ledger stores the transaction records publicly with security. Every transaction is digitally signed before transmitting to the network. So, the transactions are much more secured that shows its authenticity and integrity of the Hyperledger. The following points can be considered for the blockchain:

i) A blockchain is a decentralized, public, distributed and secure database among IoT nodes.
ii) Each IoT node has the ability to validate the blocks.
iii) Sometimes some IoT nodes (Miners) are considered as a controller in the block chains.
iv) The peer-to-peer topology is used in blockchains.
v) The hyper ledger stores all recorded transactions. The blockchains update after recorded.
vi) BT is always on. No any IoT node off the BT.

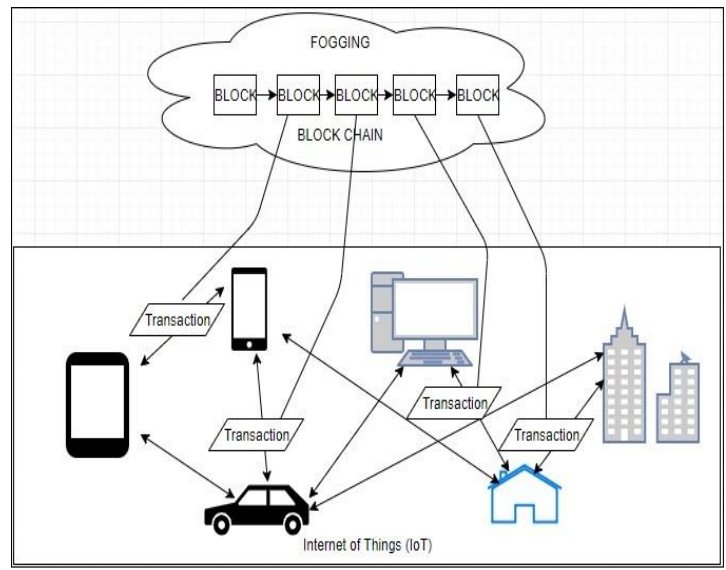

Figure 2: IoT-Fog

The above are the positive points about blockchains but there are a lot of challenges such as:
i) the processing speed of IoT device and BT
ii) Time durability
iii) storage the big data
iv) real-time connectivity
v) IoT device Intelligence power
vi) and so on…

The aim of this research is to introduce the new area of research where the top emerging technologies combined together and form a group for the ultimate goal of our proposal. It enhances the area of IoT and provides secure and authentic communication among the physical things.

The rest of the paper is organized into different sections. Section 1 introduce the research, section 2 represents the literature review, section 3 represents the proposed study and results of the proposed study and section 4 represents the conclusion of the proposed research.

## 2.0 Literature review

Fog Computing is also represented as Fogging that is a term created by Cisco. The role of fogging is to extend the cloud computing into the edge computing. The fogging is supported to the IoT network that is the main idea to develop this term. The Fogging has improved the accuracy of the system and the quality of the services. It can support the physical things that are connected together. In 2015, Cisco represented fogging for the cloud computing in the edge of the network to process the information in a more accurate way. It is used to reduce the workload of cloud computing [3]. Flavio Bonomi et. al. was presented the role of fogging and its characteristics in the internet of things [4]. Fog computing is used to minimize the information transmitted to the cloud to process, analyze and store. It is also improved the efficiency and security of the framework. In 2017, Jianbing Ni has presented an article on secure fogging for IoT. In this paper, the authors were focused on security and privacy issues faced in the communication among IoT nodes with fogging [5]. In 2017, Joy Dutta and Sarbani Roy have published an article [6] on IoT-Fog framework for smart cities. In this paper, they presented the smart building structure with the use of fog and IoT networks. In 2017, Muhtasim M., et. al. was published the thesis report [7], they have presented the security of transmitted data transactions in the internet of things network by using the blockchains technology. In the article [8], various consensus algorithms discussed and compared. They compared the algorithms such as Proof-of-Work (PoW), Proof-of-Stake (PoS), Delegated Proof-of-Stake (DPoS), Leased Proof-Of-Stake (LPoS), Proof of Elapsed Time (PET), Practical Byzantine Fault Tolerance (PBFT), Simplified Byzantine Fault Tolerance (SBFT), Delegated Byzantine Fault Tolerance (DBFT), Directed Acyclic Graphs (DAG), Proof-of-Activity (PoA), Proof-of-Importance (PoI), Proof-of-Capacity (PoC), Proof-of-Burn (PoB), and Proof-of-Weight (PoW) [9].

Table 1: Comparison of the Consensus Algorithms [9,10]

| Algorithms | Blockchain Platform | Launched Year | Languages | Smart Contracts | Advantages | disadvantages |
|---|---|---|---|---|---|---|
| PoW | Bitcoin | 2009 | C++ | No | 1. Minimize the attacks up to 50% or less<br>2. improve security | 1. More power consumption<br>2. centralized Miners |
| PoS | NXT | 2013 | Java | Yes | 1. Energy efficient<br>2. More decentralized | 1. Nothing-at-stake problem |
| DPoS | Lisk | 2016 | JavaScript | No | 1. Energy efficient<br>2. Scalable<br>3. Increased security | 1. Partially centralized<br>2. Double spend attack |
| LPoS | Waves | 2016 | Scala | Yes | 1. Fair usage<br>2. Lease Coins | 1. Decentralization Issue |
| PoET | Hyperledger Sawtooth | 2018 | Python, JavaScript, Go, C++, Java, and Rust | Yes | 1. Cheap participation | 1. Need for specialized hardware<br>2. Not good for Public Blockchain |
| PBFT | Hyperledger Fabric | 2015 | JavaScript, Python, Java REST and Go | Yes | 1. No Need for Confirmation | 1. Communication Gap<br>2. Sybil Attack |

| | | | | | 2. Reduction in Energy | |
|---|---|---|---|---|---|---|
| SBFT | Chain | 2014 | Java, Node, and Ruby | No | 1. Good Security 2. Signature Validation | 1. Not for Public Blockchain |
| DBFT | NEO | 2016 | Python, .NET, Java, C++, C, Go, Kotlin, JavaScript | Yes | 1. Scalable 2. Fast | 1. Conflictions in the Chain |
| DAG | IOTA | 2015 | Javascript, Rust, Java Go, and C++ | In Process | 1. Low-cost network 2. Scalability | 1. Implementation gaps 2. Not suited for smart contracts |
| POA | Decred | 2016 | Go | Yes | 1. Reduces the probability of the 51% attack 2. Equal contribution | 1. Greater energy consumption 2. Double signing |
| PoI | NEM | 2015 | Java, C++XEM | Yes | 1. Vesting 2. Transaction partnership | 1. Decentralization Issue |
| PoC | Burstcoin | 2014 | Java | Yes | 1. Cheap 2. Efficient 3. Distributed | 1. Favoring bigger fishes 2. Decentralization issue |
| PoB | Slimcoin | 2014 | Python, C++, Shell, JavaScript | No | 1. Preservation of the network | 1. Not for short term investors 2. Wasting coins |
| PoWeight | Filecoin | 2017 | SNARK/STARK | Yes | 1. Scalable 2. Customizable | 1. Issue with Incentivization |

## 3.0 Proposed Framework and results

The proposed framework has represented the use of Fog computing with IoT devices on the edge of the network using the blockchain technology to connect, transfer and exchange information among the IoT nodes. The transactions in the proposed framework are transmitted in the point-to-point network topology. In the network, there are some special IoT nodes called Miners. They are generally used to verify the transactions in the network. If the transactions are verified then it converts into the block and added in the blockchains that previously exist and relayed to the network. The miners play the important role to adjust the newly created block in the blockchain.

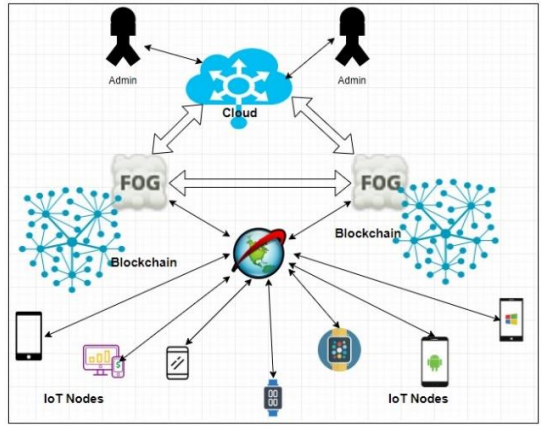

Figure 3: IoT-Fog middleware architecture using Blockchain

In this research, we have evaluated the framework using several tests. The blockchains are implemented using Hyperledger IROHA tool. The docker and docker-compose are installed on the machine. The main parameters are described in table 1 [11]. Icov coverage testing tool is used to evaluate the IoT device coverage in the range of the network.

Table 2. Main parameters for performance evaluation.

| Parameter | Possible values | Default | Description |
| --- | --- | --- | --- |
| COVERAGE | ON/OFF | OFF | Enables or disables lcov setting for code coverage generation |
| BENCHMARKING | | OFF | Enables or disables build of the Google Benchmarks library |
| TESTING | | ON | Enables or disables build of the tests |
| SWIG_PYTHON | | OFF | Enables or disables the library building and Python bindings |
| SWIG_JAVA | | OFF | Enables or disables the library building and Java bindings |

The Hyperledger IROHA tool is included many facilities such as distributed Hyperledger, Proof of Work (PoW) algorithms, P2P network, etc., Sumeragi in Hyperledger IROHA algorithm is implemented to rum blockchains. The Android and iOS packages in IROHA provide the facility to interact the IoT nodes to the blockchain. According to the Sumeragi algorithm, the IoT nodes request the transaction, and follow the following steps:

Step 1. Broadcasting: the leaders verify, orders and sign the transaction and transmit to the network.

Step 2: Verification and signing: It verifies, orders and sign the transaction and broadcasting to the authorized IoT node of the peer to peer network.

Step 3: Committed: Commit after signing.

Figure 4: Sumeragi in Hyperledger IROHA algorithm [12]

In case of server failure, the algorithm adds one another step called error control. For controlling the errors, the algorithm works with the additional server to control the errors.

Figure 5: Sumeragi in Hyperledger IROHA algorithm with error control

We have evaluated the performance of 10, 50 and 100 IoT nodes using the different parameters in the proposed framework.

Table 3: Performance of 10 IoT-nodes against searching, Examine and Selecting

|  | IoT-Node-1 ON | IoT-Node-1 OFF | IoT-Node-3 ON | IoT-Node-3 OFF | IoT-Node-5 ON | IoT-Node-5 OFF | IoT-Node-8 ON | IoT-Node-8 OFF | IoT-Node-10 ON | IoT-Node-10 OFF |
|---|---|---|---|---|---|---|---|---|---|---|
| Searching | 0.11 | 0.22 | 0.12 | 0.18 | 0.08 | 0.15 | 0.12 | 0.23 | 0.14 | 0.21 |
| Examine | 0.01 | - | 0.01 | - | 0.01 | - | 0.01 | - | 0.01 | - |
| Selecting | 0.01 | 0.01 | 0.01 | 0.01 | 0.01 | 0.01 | 0.01 | 0.01 | 0.01 | 0.01 |

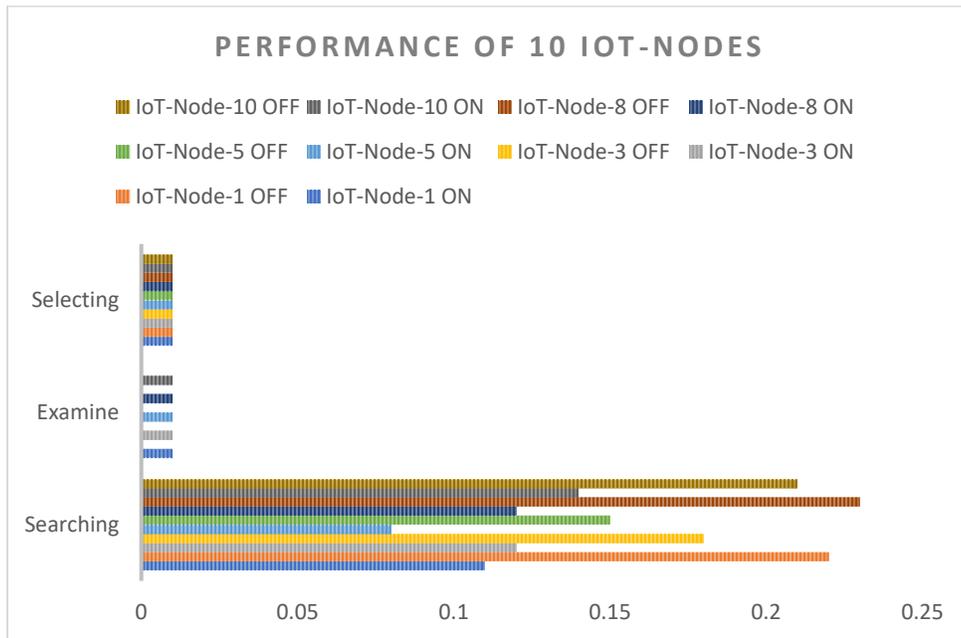

Figure 6: Performance of 10 IoT-nodes

Table 4: Performance of 50 IoT-nodes against searching, Examine and Selecting

|  | IoT-Node-1 ON | IoT-Node-1 OFF | IoT-Node-3 ON | IoT-Node-3 OFF | IoT-Node-5 ON | IoT-Node-5 OFF | IoT-Node-8 ON | IoT-Node-8 OFF | IoT-Node-10 ON | IoT-Node-10 OFF |
|---|---|---|---|---|---|---|---|---|---|---|
| Searching | 0.12 | 0.23 | 0.10 | 0.21 | 0.12 | 0.22 | 0.08 | 0.18 | 0.20 | 0.16 |
| Examine | 0.01 | - | 0.02 | - | 0.02 | - | 0.02 | - | 0.03 | - |
| Selecting | 0.02 | 0.02 | 0.02 | 0.02 | 0.01 | 0.01 | 0.02 | 0.02 | 0.04 | 0.04 |

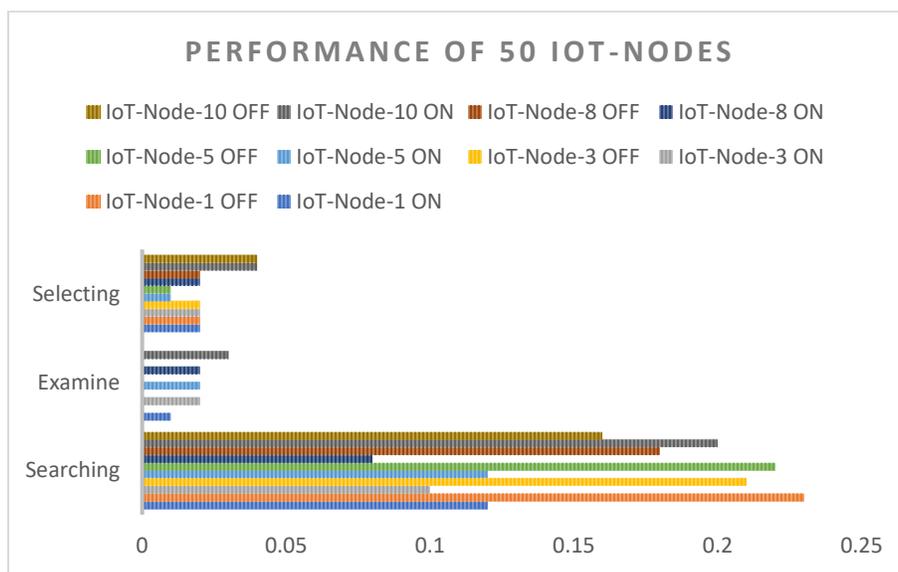

Figure 7: Performance of 50 IoT-nodes

Table 5: Performance of 100 IoT-nodes against searching, Examine and Selecting

|  | IoT-Node-1 ON | IoT-Node-1 OFF | IoT-Node-3 ON | IoT-Node-3 OFF | IoT-Node-5 ON | IoT-Node-5 OFF | IoT-Node-8 ON | IoT-Node-8 OFF | IoT-Node-10 ON | IoT-Node-10 OFF |
|---|---|---|---|---|---|---|---|---|---|---|
| Searching | 1.81 | 2.25 | 1.97 | 2.54 | 2.2 | 3.6 | 3.1 | 4.5 | 4.8 | 2.9 |
| Examine | 0.53 | - | 0.81 | - | 0.72 | - | 1.13 | - | 1.22 | - |
| Selecting | 0.24 | 0.25 | 0.37 | 0.56 | 0.29 | 0.47 | 0.38 | 0.5 | 0.55 | 0.45 |

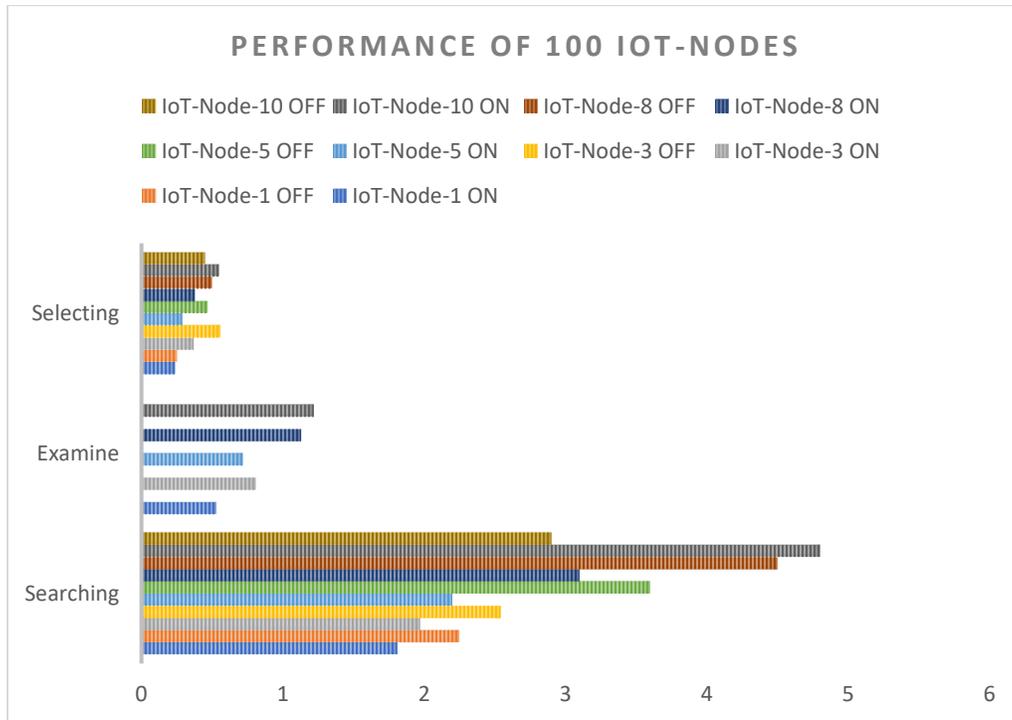

Figure 8: Performance of 100 IoT-nodes

### 4.0 Conclusion

The proposed framework acted as a combination of the Internet of Things (IoT) and fog computing. The Blockchain is used to create a hyper-distributed public authentic ledger to record the transactions. The research opened a new opportunity in this area. The framework is implemented using a different set of IoT nodes and tested. The results are found positive.

### Acknowledgment

This research is supported and funded by Deanship of scientific research, Islamic University of Madinah, Saudi Arabia (Grant No. 10/40).